\documentclass[12pt,preprint]{aastex}

\def\hii{{\rm H}{\scriptsize{\rm II}}}

\shorttitle{IC 225: a dwarf elliptical galaxy with a peculiar blue core}
\shortauthors{Q. Gu et al.}

\begin{document}

\title{IC 225: a dwarf elliptical galaxy with a peculiar blue core}

\author{Qiusheng Gu, Yinghe Zhao, Lei Shi, Zhixin Peng and Xinlian Luo}

\affil{Department of Astronomy, Nanjing University, Nanjing 210093, China}

\email{qsgu,yhzhao,xlluo@nju.edu.cn}

\begin{abstract}
   We present the discovery of a peculiar blue core in the
   elliptical galaxy IC 225 by using images and spectrum from the
   Sloan Digital Sky Survey (SDSS). The outer parts of the surface
   brightness profiles of u-, g-, r-, i- and z-band SDSS images
   for IC 225 are well fitted with an exponential function. The
   fitting results show that IC 225 follows the same relations
   between the magnitude, scale length and central surface
   brightness for dwarf elliptical galaxies. Its absolute blue
   magnitude (M$_{\rm B}$) is $-$ 17.14 mag, all of which suggest
   that IC 225 is a typical dwarf elliptical galaxy. The g$-$r
   color profile indicates a very blue core with a radius of 2
   arcseconds, which is also clearly seen in the RGB image made of
   g-, r- and i-band SDSS images. The SDSS optical spectrum
   exhibits strong and very narrow nebular emission lines. The
   metal abundances derived by the standard methods, which are
   12+log(O/H) = 8.98, log(N/O) = -0.77 and 12+log(S$^+$/H$^+$) =
   6.76, turn out to be significantly higher than that predicted
   by the well-known luminosity-metallicity relation. After
   carefully inspecting the central region of IC 225, we find that
   there are two distinct nuclei, separated by 1.4 arcseconds, the
   off-nucleated one is even bluer than the nucleus of IC 225. The
   asymmetric line profiles of higher-order Balmer lines indicate
   that the emission lines are bluer shifted relative to the
   absorption lines, suggesting that the line emission arises from
   the off-center core, whose nature is a metal-rich \hii \
   region. To the best of our knowledge, it is the first
   high-metallicity \hii \ region detected in a dwarf elliptical
   galaxy.

\end{abstract}


\section{Introduction}

   Dwarf elliptical galaxies, which typically have absolute blue
   magnitudes fainter than M$_{\rm B} = -$18 mag, are the most
   numerous type of galaxies in the local universe.
   Their formation and evolution have important
   consequences for both the galaxy luminosity function and
   constraints of cosmological models (see the
   review by Ferguson \& Binggeli 1994 and references therein).
   Recently Graham (2005) suggested that dwarf elliptical galaxies
   form a continuous extension, both chemically
   and dynamically, with the more luminous (ordinary) elliptical
   galaxies (see also Guzman et al. 2003). Bender et al.
   (1992) proposed that dwarf ellipticals could be bulges of failed disk
   galaxies which could not acquire material enough to form a
   significant disk component due to the tidal interaction by a
   nearby massive galaxy. Alternatively, Gerola et al. (1983) and
   Mirabel et al. (1992) suggested that dwarf ellipticals form as debris from
   either the explosion during the initial starburst phase of
   massive galaxies or giant-galaxy collisions.

   IC 225 is a nearby elliptical galaxy in the RC3 catalog (de
   Vaucouleurs et al. 1991), also detected in both the Markarian
   (Mrk 1038, Markarian et al. 1977) and KISO (KUG 0223+009,
   Takase \& Miyauchi-Isobe 1986) surveys of galaxies with UV
   excess. Its redshift is 0.00512 from the Sloan Digital Sky
   Survey Data Release 2 (SDSS DR2, Abazajian et al. 2004), which
   corresponds to a distance of 20.6 Mpc (for a Hubble constant of
   H$_0$ = 75 km s$^{-1}$ Mpc$^{-1}$, $\Omega_M$ = 0.3 and
   $\Omega_\Lambda$ = 0.7), the absolute blue magnitude (M$_{\rm
   B}$) is then estimated to be $-$17.14 mag. The fact that there
   is no known companion to this object within 30 arcmin by using
   the NASA/IPAC Extragalactic Database (NED) suggests that it is
   most probably isolated. Therefore, it should be classified as
   an isolated, dwarf elliptical galaxy (dE). Maehara et al.
   (1987) and Augarde et al. (1994) have shown that its nucleus
   exhibits emission lines. Based on the emission line ratios,
   Comte et al. (1994) determined that the line emission is
   dominated by photoionization from young stars and that the
   oxygen abundance is [12+log(O/H)] = 8.90.  Though the spectrum
   clearly indicates the presence of recent star-forming activity,
   it is interesting to note that 21cm HI observations show that
   there is little HI gas in IC 225 (Thuan et al. 1999; Salzer et
   al. 2002). Such active star-forming dEs were thought to be rare
   as they had lost their gas and dust long ago (Mori et al.
   1997). However, recently Drinkwater et al. (2001) discovered
   H$\alpha$ emission in about 25\% \ of the dEs during the
   spectroscopic survey of the Fornax cluster (see also Michielsen
   et al. 2004).

   In this paper, we present the imaging and spectroscopic data
   from the Sloan Digital Sky Survey for the galaxy IC 225, where
   we detect a compact blue core at the center of IC 225. More
   interestingly, we find that there are two distinct nuclei in
   the center of IC 225. The SDSS spectrum exhibits strong and
   narrow emission lines and is very similar to a metal-rich \hii\
   region. This paper is organized as follows: in Section 2 we
   present the images and optical spectrum for IC 225, both of
   which are taken out from SDSS. In Section 3 we present the results of
   surface brightness distributions for five-band SDSS images, we also
   estimate the star formation rates (SFRs) from H$\alpha$ and far-infrared
   luminosities and derive metal abundances for IC 225. Finally we discuss
   our results in Section 4 and draw conclusions in Section 5.


\section{The Data}

  The Sloan Digital Sky Survey (SDSS) is the most ambitious
  astronomical, both photometric and spectroscopic, survey project
  which has ever been undertaken (Gunn et al. 1998; Blanton et al.
  2003). Recently we cross-correlate the SDSS spectroscopic
  archive data base (Data Release 2; Abazajian et al. 2004) with
  the Third Reference Catalogue of Bright Galaxies (RC3; de
  Vaucouleurs et al. 1991), derive a sample of 1049 galaxies
  available with both morphological classification and
  spectroscopic information. This sample is large enough to study
  the circumnuclear star forming activity along the Hubble
  sequence, which will be presented by Shi et al. (2005).

  Here we present the study of low-luminosity elliptical galaxy IC
  225 by using SDSS images and spectroscopic data. Figure 1 shows
  the RGB false-color image of the central 2.0 $\times$ 2.0
  arcminutes for IC 225, taken from the SDSS DR3 Finding Chart
  Tool webpage\footnote{
  http://cas.sdss.org/dr3/en/tools/chart/chart.asp}, which
  combines information from g-, r- and i-band SDSS images by using
  the algorithm given by Lupton et al. (2004). It is clearly seen
  that there exists a blue core at the center of IC 225, while the
  light distribution is very smooth and there is no any sign of
  bar, spiral arms or any recent interacting remnants and tidal
  tails.

  In Figure 2 we show the SDSS optical spectrum for IC 225. The
  SDSS fiber size is 3 arcseconds, which corresponds to the
  diameter of 300pc at the distance of IC 225.  We can easily
  identify strong nebular emission lines, such as Balmer lines
  (H$\alpha$, H$\beta$ and H$\gamma$), [OIII]$\lambda$4959,5007, \
  [NII]$\lambda$6548,6583 \ and [SII]$\lambda$6716,6732, etc, as
  already shown by Augarde et al. (1994). Two interesting points
  are that: 1) all emission lines are very narrow, the full widths
  at half maximum (FWHM) of Balmer emission lines are just as
  narrow as those of forbidden lines, such as [OIII] and [SII],
  which are {\it only} 170 km $s^{-1}$; 2) higher-order Balmer
  absorption lines in the wavelength range of 3800 $-$ 4000 \AA \
  are clearly presented, which have been taken as the unambiguous
  evidence of intermediate-aged ($\sim 10^8$ yr) stellar
  populations (Gonzalez Delgado, Leitherer \& Heckman 1999).

\section{Results}

\subsection{Radial Profiles}

 It is known that the surface brightness of dwarf elliptical galaxies
 could be better described by
 an exponential function (Faber \& Lin 1983; Binggeli, Sandange \&
 Tarenghi 1984; Graham \& Guzman 2003).  In Figure
 3 we show the surface brightness distribution for SDSS 5-band (u,
 \ g, \ r, \ i, and \ z) images by using the standard tasks in {\it
 IRAF} \footnote{IRAF is distributed by the National Optical
 Astronomy Observatories, which is operated by the Association of
 Universities for Research in Astronomy, Inc., under cooperative
 agreement  with the National Science Foundation.}.

 The central 4-5 arcseconds in IC 225 are clearly affected by the
 presence of recent active star forming activity (as can be seen in the color profile
 shown in Fig. 4), we thus only fit the surface brightness profiles beyond 5
 arcseconds for all SDSS 5-band images by using both
 a de Vaucouleurs $R^{1/4}$ and an exponential law. We find that a single
 exponential law gives a statistically better fit (e.g. the minimum $\chi^2$),
 whose $\chi^2$ is typically 50\% smaller than the fit by the de Vaucouleurs $R^{1/4}$ law.
 In Fig. 3 we also show the best fits by an exponential law as the solid lines and
 the fitting results are summarized in Table 1, where we present
 the scale length (R$_{\rm S}$), central surface brightness ($\mu_0$)
 and the rms values of the fitting. We find that the extrapolated central surface
 brightness and the scale length for IC 225 obey the same relations between
 absolute magnitude and central surface brightness/scale length derived for a large sample
 of dwarf elliptical galaxies (e.g. Fig. 9 Binggeli \& Cameron 1991), suggesting that IC 225
 is a typical dwarf elliptical galaxy.

 After carefully matching PSF profiles of g- and r-band images, we also
 measure the g$-$r color distribution for IC 225 in Figure
 4. It is very interesting to find that the g$-$r color is
 exceptionally bluer in the center 2" region than that of the
 outer part, just as the blue cores found in the higher redshift
 elliptical galaxies (Menanteau et al. 2001). In the central
 region of IC 225, the mean color of g$-$r is 0.22 mag,
 while in the outer region, the g$-$r color distribution is nearly
 uniform but much redder (0.54 mag).
 The radial distribution of color (g$-$r)
 clearly indicates that the starburst activity in IC 225 is highly
 concentrated in the central region and also confirms the finding
 of the blue core in the false-color RGB image.

\subsection{Star Forming Activity}

 An important and unavoidable issue in the study of emission-line
 spectra is the dilution from the underlying old stellar
 populations, particularly for the Balmer lines. One of popular
 methods to remove the contribution from old stellar population is
 to use stellar population synthesis model (Kauffmann et al. 2003;
 Cid Fernandes et al. 2004; Hao et al. 2005). We use the same
 stellar population synthesis code, {\scriptsize STARLIGHT
 version 2.0}, as Cid Fernandes et al. (2004) to study the stellar
 properties of IC 225 through fitting the SDSS optical spectrum.
 The best fit is shown in Figure 5, where we plot the synthetical
 and pure emission-line (subtracting the best matched model from
 the observed spectrum) spectra, together with flux- and mass-
 fractions of 15 different-aged Simple Stellar Populations (SSP)
 which are chosen from Bruzual \& Charlot (2003). The light
 contributions from young (age $<$ 10$^8$ yr), intermediate-aged
 (10$^8$ $\le$ age $<$ 10$^9$ yr) and old ($\ge$ 10$^9$ yr) stellar
 components are 40\%, 56\% and 4\%, respectively.

   On the other hand, we can also use the equivalent widths (EWs)
   of the Balmer emission lines to constrain the ages of the
   underlying (ionizing) stellar populations (e.g. Leitherer \&
   Heckman 1995; Stasi\'nska \& Leitherer 1996). For IC 225, the
   EWs of H$\alpha$ and H$\beta$ emission lines relative to the
   continuum due to the young (age $<$ 10$^8$ yr) stellar
   population alone are 227.5 and 45 \AA, respectively, which
   suggest the age of young starburst to be 6 $\sim$ 7 $\times
   10^6$ yrs as compared with a Simple Stellar Population model by
   adopting Salpeter's Initial Mass Function (IMF) with $\alpha =
   2.35$ and $\rm M_{up} = 100 M\odot$ at solar metallicity
   (Stasi\'nska \& Leitherer 1996).

 From the pure emission-line spectrum, we can measure the accurate
 fluxes of emission lines as listed in Table 2, the line flux errors are
 typically less than 5\%. Using emission
 line flux ratios, the so-called BPT diagrams (Baldwin, Phillips
 \& Terlevich 1981; Veilleux \& Osterbrock 1987), we confirm that
 the nebular photoionization in IC 225 is dominated by starburst
 activity (Comte et al. 1994).

 By assuming Case B recombination and a standard reddening law
 (Cardelli, Clayton \& Mathis 1989), we could estimate the nebular
 extinction from the observed $H\alpha/H\beta$ \ and
 $H\gamma/H\beta$ \ Balmer decrements (see Torres-Peimbert,
 Peimbert \& Fierro 1989), which are

 \begin{equation}
 A_{V}^{H\alpha} = 6.60 \times \log
(\frac{F_{H\alpha}/F_{H\beta}}{I_{H\alpha}/I_{H\beta}})
 \end{equation}

\begin{equation}
 A_{V}^{H\gamma} = -16.37 \times \log
(\frac{F_{H\gamma}/F_{H\beta}}{I_{H\gamma}/I_{H\beta}})
 \end{equation}

 \noindent Where $F_{H\alpha}/F_{H\beta}$,$F_{H\gamma}/F_{H\beta}$
 and $I_{H\alpha}/I_{H\beta}$, $I_{H\gamma}/I_{H\beta}$ are the
 observed and intrinsic Balmer decrements, respectively. In this
 paper, we adopt the intrinsic ratios of $I_{H\alpha}/I_{H\beta}$
 and $I_{H\gamma}/I_{H\beta}$ to be 2.87 and 0.466, respectively
 (Osterbrock 1989). For IC 225, the observed
 $F_{H\alpha}/F_{H\beta}$ and $F_{H\gamma}/F_{H\beta}$ (measured
 from the pure emission line spectrum) is equal to 3.78 and 0.419,
 then the nebular extinction, $A_{V}^{H\alpha}$ and
 $A_{V}^{H\gamma}$, are estimated to be 0.79 and 0.76 mag,
 respectively. These two values are nearly the same, which
 indicates that the stellar population synthesis model is
 successful to remove the old stellar contribution, the fluxes of
 emission features are measured accurately and reliably. In the
 following we will assume the nebular extinction ($A_{V}$) to be
 0.775 to derive the extinction-corrected fluxes for emission lines.

 By using the empirical relation between nebular emission line luminosity
 and star formation rate (SFR) given by Kennicutt (1998), we can
 derive the SFR from H$\alpha$ luminosity. For IC 225, the
 extinction-corrected H$\alpha$ luminosity is $L_{\rm H\alpha} =
 3.0 \times 10^{39} \ erg s^{-1}$ and the corresponding SFR$_{\rm
 H\alpha}$ for the central 300 pc in IC 225 is equal to 0.024
 M$\odot$ yr$^{-1}$. At the same time, we could also estimate the
 SFR from the far infrared (FIR) luminosity (also given by
 Kennicutt 1998). For IC 225, the IRAS fluxes at 12$\mu$m,
 25$\mu$m, 60$\mu$m and 100$\mu$m are 0.085, 0.142, 0.192 and
 0.477 Jy, respectively, which are taken from the IRAS Faint
 Source Catalog (Moshir et al. 1989). Thus, the FIR luminosity is
 equal to 1.55 $\times$ 10$^8$ L$\odot$ and the corresponding
 SFR$_{\rm FIR}$ is equal to 0.027 M$\odot$ yr$^{-1}$, nearly the
 same as that derived from H$\alpha$ luminosity. As we know, IRAS
 aperture size is rather large, typically a few arcminutes.  However,
 the SFR derived from H$\alpha$ emission could be underestimated
 because of SDSS aperture effects, which will be addressed in the
 next Section where we present a detailed description of these effects.

 \subsection{Metal Abundances}

   Due to low redshift of IC 225,
   [OII]$\lambda$3727 does not appear in the SDSS spectrum. We find
   that Augarde et al. (1994) presented the relative intensities
   of emission lines: [OII]$\lambda$3727, H$\alpha$,
   [NII]$\lambda$6584 and [SII]$\lambda$6725 to H$\beta$ line
   flux, which are 1.448, 3.804, 0.823, and 1.04, respectively,
   while our measurements of relative intensities for H$\alpha$,
   [NII]$\lambda$6584 and [SII]$\lambda$6725 are 3.782, 0.836, and
   1.155, respectively. The differences are less than 10\%, so we
   just simply assume [OII]$\lambda$3727/H$\beta$ = 1.448 for the
   further analysis.

   Following the standard methods to compute metal abundances
   (Pagel et al. 1992) and assuming that most of the oxygen is in
   the first- and second-ionization levels, therefore O/H =
   (O$^+$+O$^{++}$)/H$^+$ and N/O = N$^+$/O$^+$, the derived ionic
   abundances are that 12+log(O/H) = 8.98, log(N/O) = -0.77 and
   12+log(S$^+$/H$^+$) = 6.76, which confirm the oxygen abundance
   of 8.90 by Comte et al. (1994) and turn out to be significantly
   higher than solar. By using the luminosity-metallicity relation
   for a sample of isolate nearby dwarf galaxies and \hii\ regions
   (Figure 2 in Duc et al. 2004), for IC 225, M$_B$ = -17.14, the
   predicted oxygen metallicity is 8.2, which is far smaller than
   the observed one(8.98), with the maximum difference of 0.78 dex.

   We note that the abundance commonly used in the luminosity-metallicity relation of
   dwarf ellipticals (e.g. Barazza \& Binggeli 2002) is the iron
   abundance in the stars. It is well known that because the iron
   production has a longer timescale (associated with SN Ia) than the
   oxygen (associated with SN II) an overabundance of oxygen with respect
   to iron is commonly found in the stellar spectra of elliptical
   galaxies (Thomas et al. 2002). This
   overabundance is expected to be even more significant when comparing
   iron in the stars with the oxygen in the ISM. Therefore, it might be
   possible that although the oxygen abundance of the ISM is slightly
   above solar, the iron abundance in the stars is not as different from
   that found in dwarf ellipticals of similar luminosity. Note, however, that the O/Fe
   overabundance seems to be smaller in low luminosity
   ellipticals than in massive ones. In the case of IC225, the low S/N for
   iron absorption lines hampers us for estimating the iron abundance. Using the
   empirical calibration for the mean (universal) luminosity $-$ metallicity relation
   given by Barazza \& Binggeli (2002), we predict the iron abundance [Fe/H]
   for IC 225 to be -0.65.

\section{Discussion}

   We have shown that for IC 225 the absolute blue magnitude is
   $-$ 17.14 mag, the outer parts of surface brightness distributions are
   well fitted by an exponential law, and the fitting results show that
   IC 225 follows the same relations between the magnitude and the scale length
   and central surface brightness for dwarf elliptical galaxies,  all of
   which suggest that IC 225 is a typical dwarf elliptical galaxy (Binggeli 1994;
   Kormendy \& Bender 1994).
   However, several observational evidences, such as strong but narrow nebular
   emission lines, the blue core in the RGB image,
   blue g$-$r color in the central 2" region, support that the young star-forming
   activity is occurring at the nuclear region.

   We search in the view field of 30 arcminutes of IC 225 by
   using the NASA/IPAC Extragalactic Database (NED), and don't
   detect any interacting companion. But when we check the
   SDSS images carefully, we find that there are two distinct
   nuclei, separated by 1.4 arcseconds in the central region of IC
   225, which is shown in Fig. 6. The off-nucleated core has very
   blue (g$-$r) color, even bluer than the nucleus of the galaxy; it
   almost disappears in the i-band but it has higher surface
   brightness than the nucleus in the u-band image. Could it be
   a dwarf galaxy or a halo cluster swallowed by IC225 and thus
   trigger the starburst activity in IC 225 ?
   The high metallicity (12+log(O/H)=8.98) rules out such possibility of
   a dwarf galaxy or a halo cluster, which is used to be metal-poor
   (1/10 solar or even less).
   However, we still can't have a direct answer due to the no spatially-resolved information
   from the SDSS spectrum, but we can have a try to probe its nature by
   measuring the recession velocity of both several emission lines and absorptions
   lines, looking for a velocity difference that could indicate that while the absorption
   comes in the most part from stars in the high-surface brightness nucleus, the emission
   probably arises from an off-nuclear HII region with slightly different radial velocity
   and by using the well-known luminosity-metallicity relation (Duc et al.
   2004; Tremonti et al. 2004).

   In order to make sure that the off-nuclear core is
   physically associated with IC 225 and is the place where the line
   emission arises from, we try the following ways. First, we find that the
   off-nucleated region in the u-band image is spatially more extended than the PSF profile that
   excluded a QSO as the background source or a field star as the foreground source.
   Second, we have measured the aperture color of the nucleus and off-nuclear region. After carefully
   checking the PSF profiles of g- and r-band images, we smooth the g-band
   image by convolving a Gaussian kernel to match the r-band PSF. By
   using tasks of {\it geomap} and {\it geotran} in IRAF, we rotate and shift
   the g-band image relative to the r-band image so that we could accurately
   measure the aperture colors (g$-$r) of the nucleus and the off-nuclear region with
   aperture size of 1.5 pixels, which are 0.23 and 0.20 mag, respectively, and quantitatively
   confirm that the off-nuclear region is even bluer than the nucleus of IC225.
   Third, fortunately the SDSS spectrum of IC 225
   shows higher-order Balmer absorption lines so that we can compare the recession velocities
   of both Balmer emission lines and absorption lines. As shown in the inset of Fig. 2, higher-order
   Balmer emission lines
   (such as H$_9$, H$_8$, H$\epsilon$ and H$\delta$) are clearly present in the centers of absorption
   lines and these emission-absorption line profiles are all asymmetric, the red wings
   are more deeper than the blue wings, which suggests that the emission lines are bluer shifted
   relative to the absorption lines. Quantitatively we use the task {\it specfit} in IRAF to fit the profiles
   of H$\epsilon$ and H$\delta$ (as shown in Fig. 7), and confirm that the centroids of emission lines
   are bluer shifted by 0.4 and 0.5 \AA \  relative to the absorption lines, respectively,
   suggesting that while the absorption comes
   in the most part from stars in the high-surface brightness nucleus the emission probably arises
   from an off-nuclear \hii \ region with slightly different radial velocity.
   To the best of our knowledge, it is the first
   high-metallicity \hii \ region detected in a dwarf elliptical
   galaxy. A similar metal-rich \hii\ region is detected in the
   spiral galaxy NGC 1232 by Castellanos, Diaz \& Terlevich (2002),
   whose metal abundances are 12+log(O/H) = 8.95, log(N/O) = -0.81
   and 12+log(S$^+$/H$^+$) = 6.67.

   In order to unambiguously understand the nature of these two
   cores, their relation and the origin of star-forming activity
   in the central region of IC 225, we need further observational
   data. Such long-slit optical spectroscopy for this galaxy is
   under consideration.

   Finally, since the size of the SDSS fiber (1.5" in radius) is comparable to the
   distance between the two regions identified near the center of IC 225,
   the actual positioning of the fiber is critical for our conclusions about star
   forming activity in IC 225. For the SDSS, the centroid of an object is defined
   as the first moment of the light distribution, the r CCDs are served as the
   astrometric reference CCDs and the relative astrometric accuracy between the r
   filter and each of the other filters (u, g, i, z) is about 25$-$35 mas (rms).
   Pier et al. (2003) have described the astrometric calibration and centroiding
   algorithm in detail and also estimated centroiding errors for galaxies, ranging
   from about 60 mas at r $\sim$ 17 mag to 170 mas at r $\sim$ 22 mag,
   which are independent of seeing. Since it is the r-band image which was
   used to determine where the fiber should be placed and the nominal coordinates
   of the fiber center, for IC 225 a significant part of the emission from the
   off-nuclear region was missing, thus it will lead to underestimate the total SFR
   from the H$\alpha$ luminosity detected within the fiber.

\section{Conclusions}

   In this paper, we present new imaging and spectroscopic data
   for IC 225 from the Sloan Digital Sky Survey (SDSS). The outer
   parts (beyond 5") of surface brightness distributions of u-,
   g-, r-, i- and z-band SDSS images for IC 225 are well fitted
   with an exponential function, the fitting results show that IC
   225 follows the same relations between the magnitude, scale
   length and central surface brightness for dwarf elliptical
   galaxies, and its absolute blue magnitude (M$_{\rm B}$) is $-$
   17.14 mag, suggesting that IC 225 is a typical dwarf elliptical
   galaxy. However,  the SDSS optical spectrum exhibits strong and
   very narrow nebular emission lines, the RGB image and g$-$r
   color distribution indicate a blue core in the central region
   of IC 225, where we find there exist two distinct nuclei,
   separated by 1.4 arcseconds, the off-nucleated core is even
   bluer than the nucleus of the galaxy. From the SDSS spectrum,
   we derive the oxygen abundance is significantly higher than
   that predicted by the metallicity-luminosity relation. At the
   same time, the asymmetric line profiles of higher-order Balmer
   lines indicate that the emission lines are bluer shifted
   relative to the absorption lines, which supports that the line
   emission comes from the off-center region, we thus propose that
   the off-nucleated core is a metal-rich \hii\ region, which is
   the first high metallicity \hii\ region known in the dwarf
   elliptical galaxies. The discovery of a metal-rich \hii\ region
   in a dwarf elliptical galaxy will provide important clues for
   the understanding of the star formation history and chemical
   evolution of dwarf galaxies and of the luminosity-metallicity
   relation.

\begin{acknowledgements}
 The authors are very grateful to the anonymous referee for his/her
 careful reading of the manuscript and thoughtful and instructive comments
 which significantly improved the content of the paper.
 We would like to thank Jorge Melnick and Jiasheng Huang for valuable
 discussions and suggestions as well as Roberto Cid Fernandes for
 sending us the updated code for stellar population synthesis.
 This work is supported by the National Natural Science Foundation
 of China under grant 10221001 and the National Key Basic Research
 Science Foundation (NKBRSG19990754). Funding for the creation and
 distribution of the SDSS Archive has been provided by the Alfred
 P. Sloan Foundation, the Participating Institutions, the National
 Aeronautics and Space Administration, the National Science
 Foundation, the U.S. Department of Energy, the Japanese
 Monbukagakusho, and the Max Planck Society. The SDSS web site is
 http://www.sdss.org/.  The SDSS is managed by the Astrophysical
 Research Consortium (ARC) for the Participating Institutions. The
 Participating Institutions are The University of Chicago,
 Fermilab, the Institute for Advanced Study, the Japan
 Participation Group, The Johns Hopkins University, Los Alamos
 National Laboratory, the Max-Planck-Institute for Astronomy
 (MPIA), the Max-Planck-Institute for Astrophysics (MPA), New
 Mexico State University, University of Pittsburgh, Princeton
 University, the United States Naval Observatory, and the
 University of Washington. This research has made use of the
 NASA/IPAC Extragalactic Database (NED) which is operated by the
 Jet Propulsion Laboratory, California Institute of Technology,
 under contract with the National Aeronautics and Space
 Administration.

\end{acknowledgements}

\clearpage
   \begin{figure}
   \centering
   \plotone{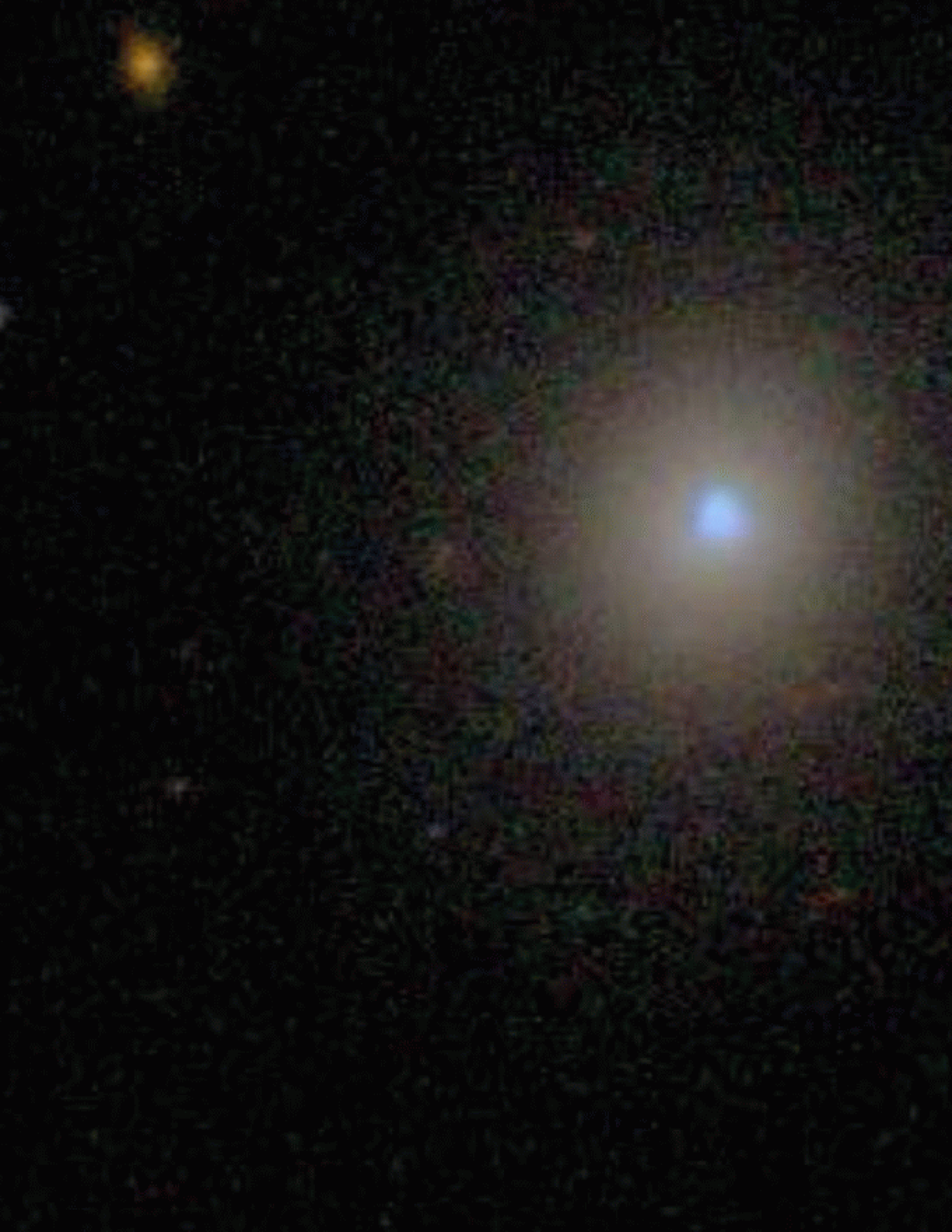}
   \caption{The RGB false-color image of IC 225, which combines
   g-, r- and i-band images from the Sloan Digital Sky Survey. A
   blue core is seen in the central region. The image size
   is 2.0 $\times$ 2.0 arcminutes. }
   \label{f1}%
    \end{figure}
\clearpage

   \begin{figure*}
   \centering
   \plotone{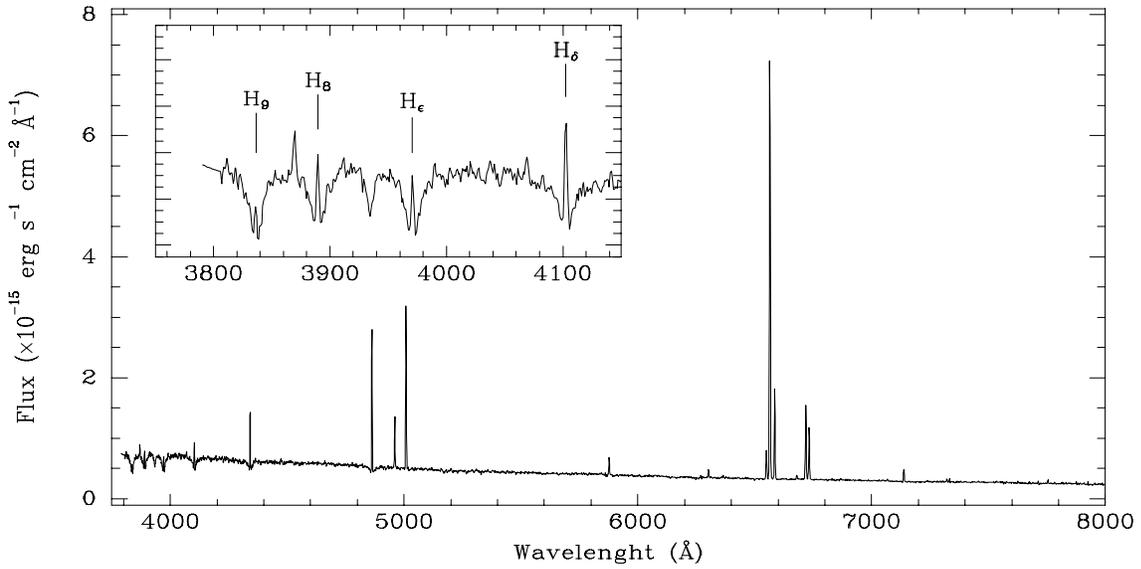}
   \caption{The optical spectrum of IC 225, taken from
    the Sloan Digital Sky Survey. The inset shows enlargement of
    the wavelength range of 3750 - 4150 \AA, where the higher-order
    Balmer absorption lines are clearly detected and these emission-absorption
    line profiles are all asymmetric.}
              \label{fig02}%
    \end{figure*}
%
\clearpage

   \begin{figure}
   \centering \plotone{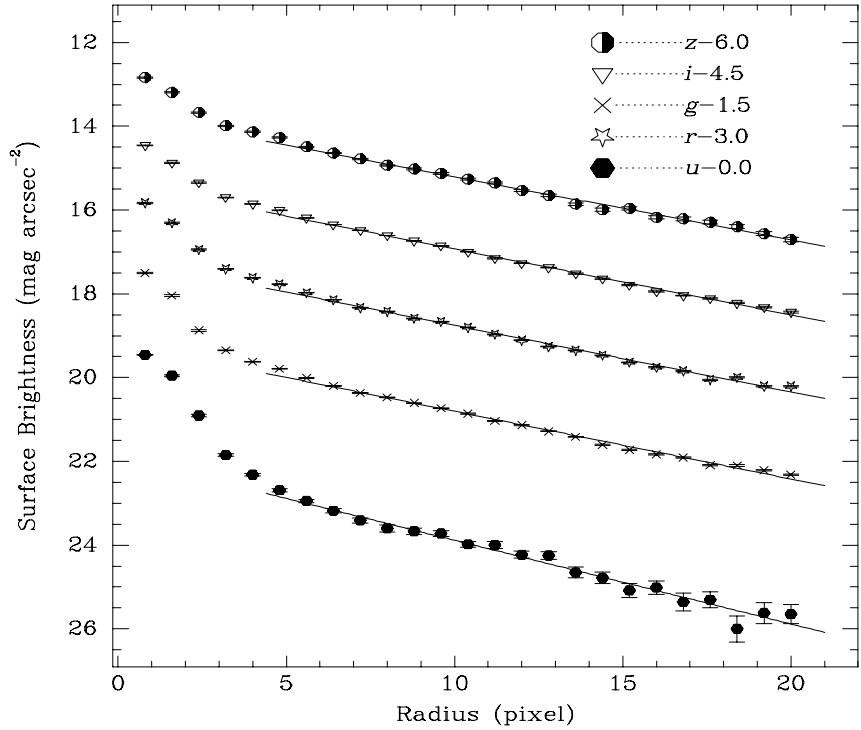} \caption{Surface brightness of IC
   225 as a function of radius for SDSS u-, g-, r-, i, and z-bands
   images. The solid lines are the best fits with an exponential law
   only for the outer parts of the surface brightness profiles.
   The profiles have been shifted for clarity, error bars are also
   shown for each measured point.  \label{fig03} }
    \end{figure}

\clearpage

   \begin{figure}
   \centering
   \plotone{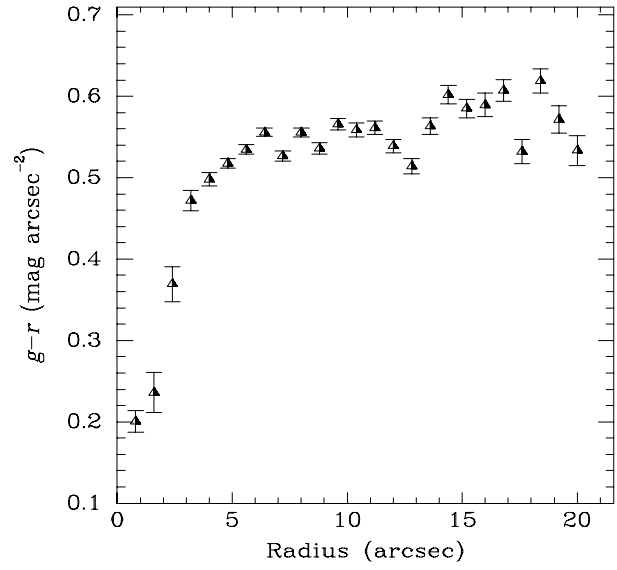}
   \caption{The g$-$r color distribution for IC 225. \label{fig04}}
    \end{figure}

\clearpage

%
   \begin{figure*}
   \centering
   \plotone{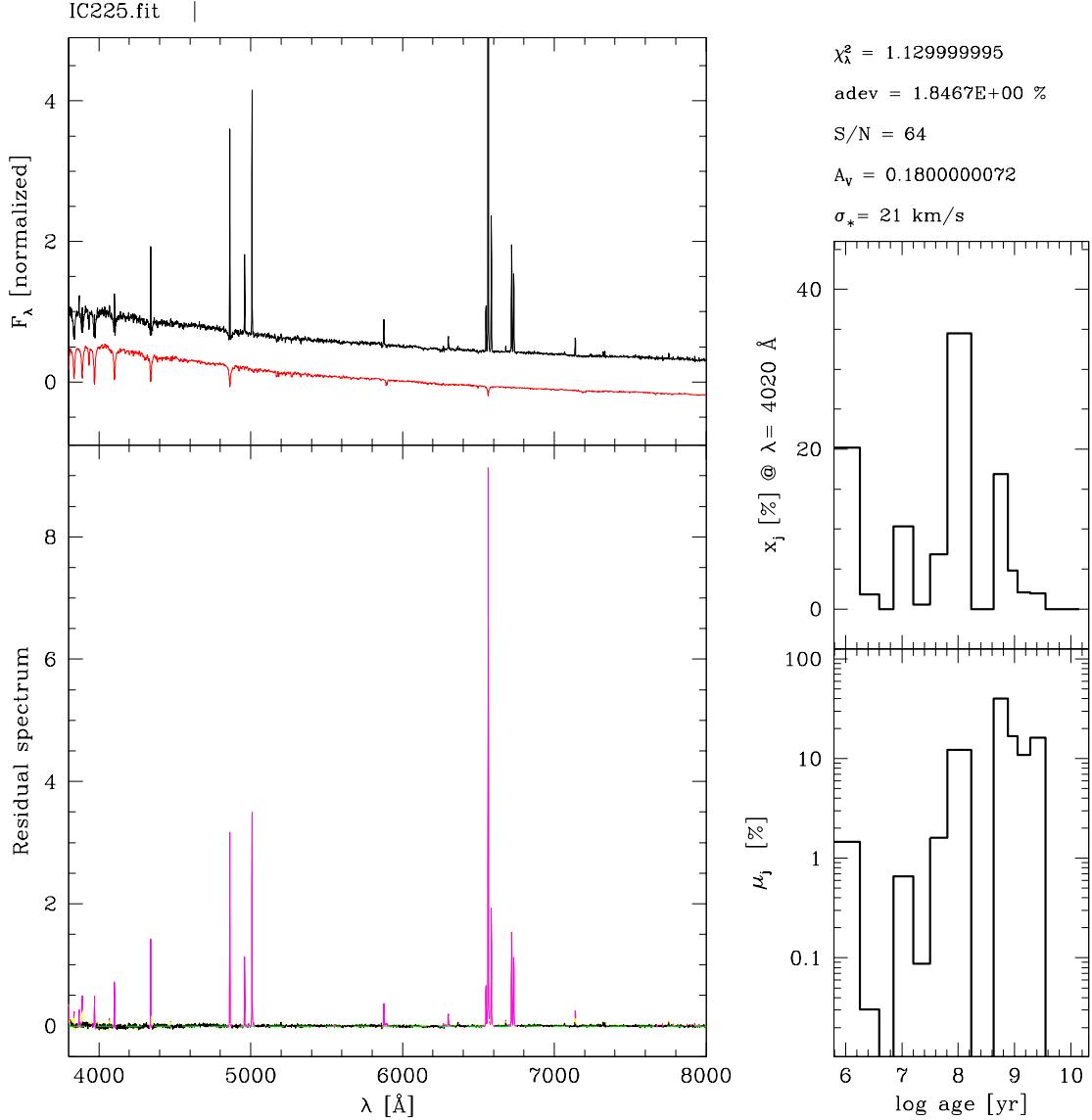}
      \caption{Results of the spectral fitting for IC 225. The top
   left panel shows the logarithm of the observed and the
   synthetic spectra, which is shown in red color and has been
   shifted by -0.5 for clarity. The $F_\lambda^O - F_\lambda^M$
   residual spectrum is shown in the bottom left panel.   Spectral
   regions actually used in the synthesis are plotted with black
   color, while masked regions are plotted with the color of
   magenta. Panels in the right   show the population vector
   binned in the 15 ages of the base. The top panel corresponds to
   the population vector in flux fraction, normalized to
   $\lambda_0 = 4540$ \AA, while the corresponding mass fractions
   vector is shown in the bottom. }
         \label{fig05}
   \end{figure*}

\clearpage

   \begin{figure*}
   \centering
   \plotone{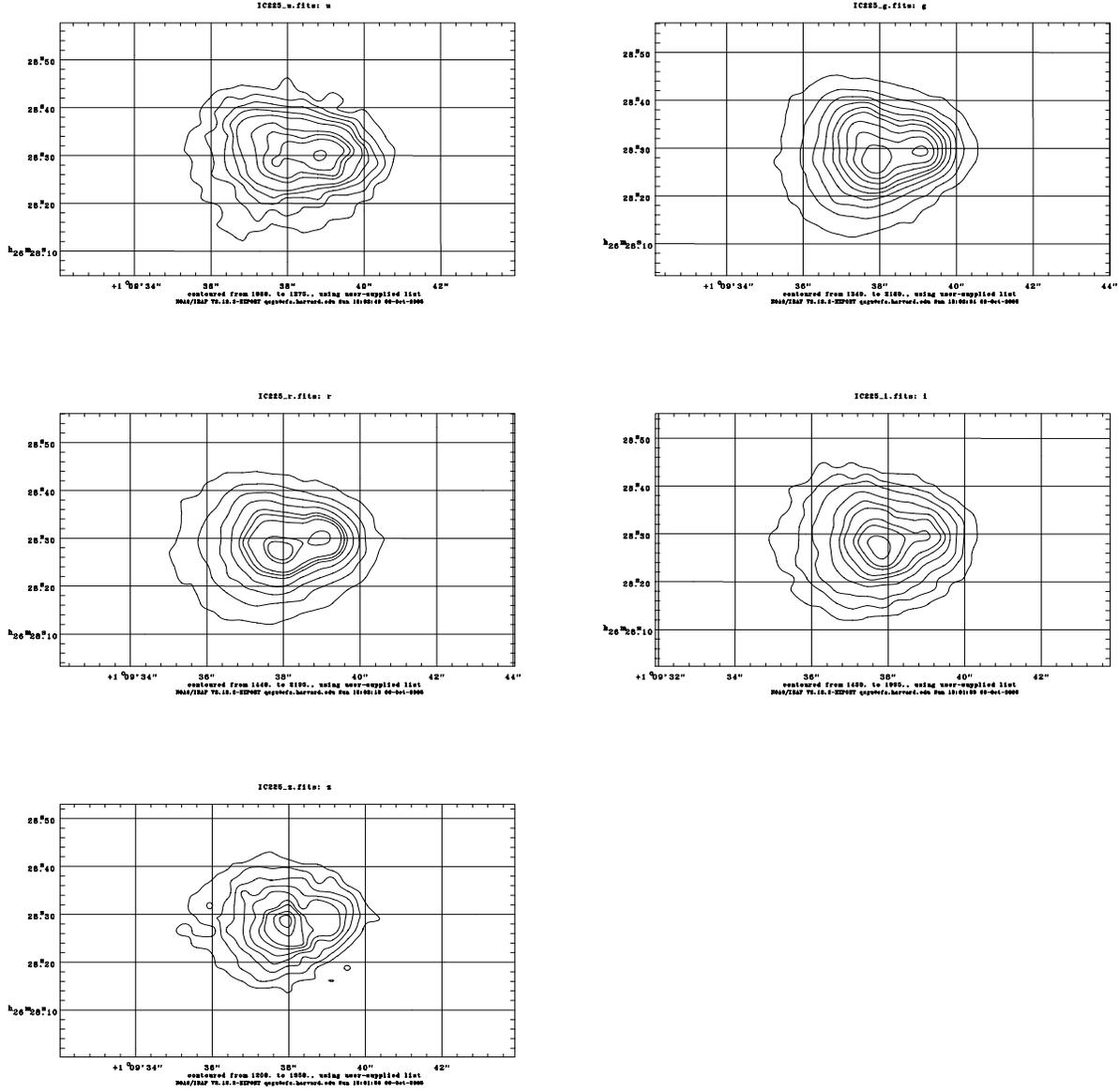}
   \caption{The u-, g-, r-, i- and
   z-band contour plots for the central region of IC 225. Two
   distinct nuclei are unambiguously found, the off-nucleated core
   is even bluer than the nucleus of the galaxy, which has higher
   surface brightness in the u-band but almost disappears in
   the i-band image. \label{fig06}}
    \end{figure*}

    \clearpage

   \begin{figure*}
   \centering
   \plotone{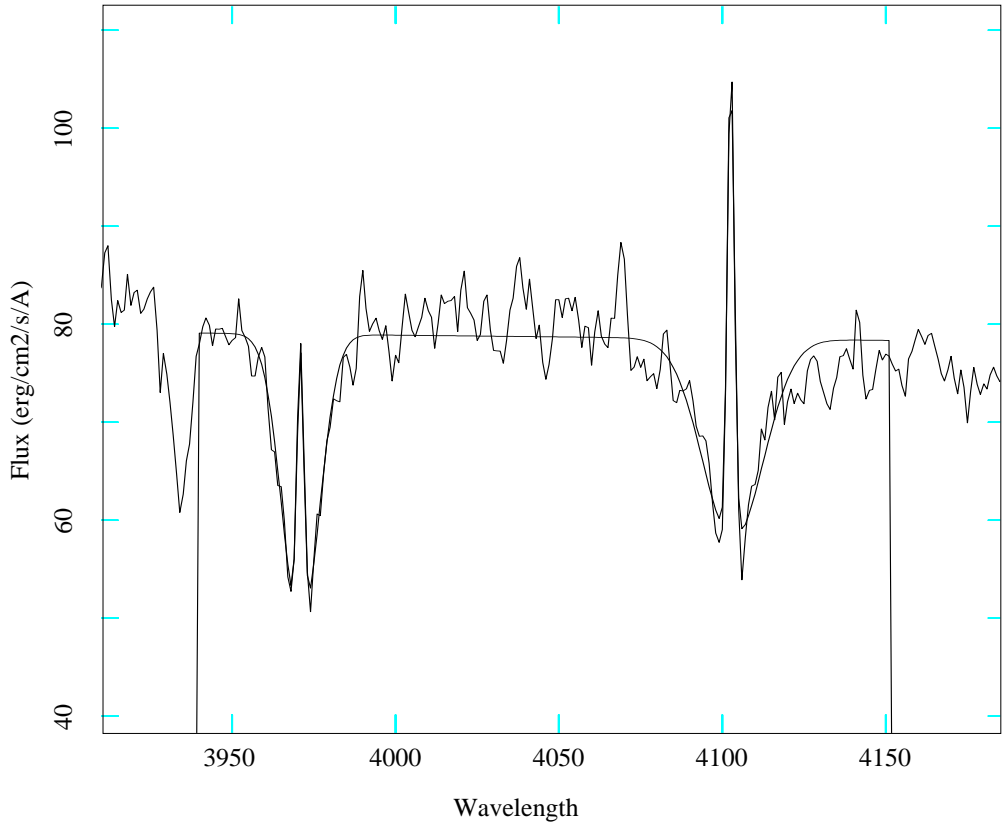}
   \caption{The spectral fitting to the absorption+emission line profiles of H$\epsilon$ and H$\delta$.
   The flux is in unit of 10$^{-17} erg cm^{-2} s^{-1} \AA^{-1}$.  \label{fig07}}
    \end{figure*}

\clearpage

\begin{table}
\begin{center}
\caption{Results of fitting the outer parts of surface brightness
distribution with an exponential law. \label{table1}}
\begin{tabular}{cccr}
\tableline\tableline
 Band  &  R$_{\rm S}$ & $\mu_{0}$ & rms \\
       &  arcsec  & mag/arcsec$^2$ &  \\
\tableline
    u     &     5.43 $\pm$ 0.26   &  21.89 $\pm$ 0.13 & 0.15 \\
    g     &     6.72 $\pm$ 0.14   &  20.69 $\pm$ 0.05 & 0.05 \\
    r     &     6.84 $\pm$ 0.15   &  20.17 $\pm$ 0.05 & 0.05 \\
    i     &     6.90 $\pm$ 0.09   &  19.86 $\pm$ 0.03 & 0.03 \\
    z     &     7.21 $\pm$ 0.16   &  19.70 $\pm$ 0.05 & 0.05 \\
\tableline
\end{tabular}
\end{center}
\end{table}

\clearpage

\begin{table}
\begin{center}
\caption{Emission lines properties for IC 225. \label{table2}}
\begin{tabular}{clr}
\tableline\tableline
 Wavelength (\AA) & Ion & flux\tablenotemark{a} \\
\tableline
        3835    & H9        &  28.9 \\
    3889    & H8        &  96.4 \\
    3970    & H$\epsilon$ & 120.2 \\
    4101    & H$\delta$   & 164.8 \\
    4340    & H$\gamma$ & 361.2 \\
    4363    & [O III]     & 12.2  \\
    4861    & H$\beta$   & 862.6 \\
    4959    & [O III]    & 331.5 \\
    5007    & [O III]    & 994.6 \\
    6300    & [O I]     & 90.2 \\
    6548    & [N II]    & 240.3 \\
    6563    & H$\alpha$ & 3262.3 \\
    6583    & [N II]    & 720.9 \\
    6716    & [S II]    & 577.5 \\
    6732    & [S II]    & 420.1 \\
\tableline
\end{tabular}

\tablenotetext{a}{fluxes are in units of 10$\rm ^{-17}$ erg \ s$^{-1}$ \ cm$^{-2}$}
\end{center}
\end{table}

\end{document}